\documentclass[aps,prb,superscriptaddress,twocolumn,longbibliography]{revtex4-2}
\usepackage{graphicx,xcolor}
\usepackage{amsmath,amssymb}
\usepackage{physics}
\usepackage[colorlinks,citecolor=blue,urlcolor=blue]{hyperref}


\begin{document}

\title{Multipolar Piezoelectricity and Anisotropic Surface Transport in Alterelectrics}

\author{Amber C. H. Visser}
\affiliation{Institute of Physics, University of Amsterdam, Science Park 904, 1098 XH Amsterdam, The Netherlands}
\affiliation{Cavendish Laboratory, University of Cambridge, JJ Thomson Avenue, Cambridge CB3 0HE, United Kingdom}

\author{Viktor K\"onye}
\affiliation{Institute of Physics, University of Amsterdam, Science Park 904, 1098 XH Amsterdam, The Netherlands}

\author{Oleg Janson}
\affiliation{Institute for Theoretical Solid State Physics, IFW Dresden, 01069 Dresden, Germany}

\author{Jeroen van den Brink}
\affiliation{Institute for Theoretical Solid State Physics, IFW Dresden, 01069 Dresden, Germany}
\affiliation{W\"urzburg-Dresden Cluster of Excellence ctd.qmat, Germany}
\affiliation{Institute  for  Theoretical  Physics,  TU  Dresden,  01069  Dresden,  Germany}

\author{Corentin Coulais}
\affiliation{Institute of Physics, University of Amsterdam, Science Park 904, 1098 XH Amsterdam, The Netherlands}

\author{Jasper van Wezel}
\email{vanwezel@uva.nl}
\affiliation{Institute of Physics, University of Amsterdam, Science Park 904, 1098 XH Amsterdam, The Netherlands}

\begin{abstract}
Alter\emph{magnets} are an emergent class of materials combining features of ferro- and antiferro-magnetic materials. They have spin-separated bands normally associated with ferromagnets, but a vanishing net magnetization. Moreover the symmetries giving rise to $d$-wave altermagnetism can provide them with a particular anisotropic, quadrupolar (i.e. with equal and opposite values when strained in perpendicular directions) piezomagnetism. Observing that the same symmetries provide a natural place to look for hyberbolic wave dispersion, this raises the question which properties are intrinsically linked to magnetism and which are determined by the symmetry. Here, we disentangle these concepts by introducing an alternative to altermagnets, based on electric polarization. These alter\emph{electrics} display quadrupolar piezoelectricity and a hyperbolic dispersion, which we demonstrate conceptually within a simplified model as well as a first-principles material realization. We furthermore establish that a counterpart of the spin-separated bands is formed by surface modes which allow for surface dependent anisotropic electronic transport analogous to the spintronic applications proposed for altermagnets.
\end{abstract}

\maketitle

\begin{figure}[b!]
\begin{center}
\includegraphics[width=0.99\columnwidth]{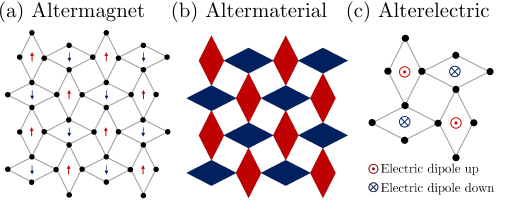}
\caption{\label{fig1} {\bf From altermagnet to alterelectric.} (a) In altermagnets, overall time-reversal symmetry is broken, while the combination of spin-rotations and a discrete lattice rotation or mirror is preserved. In this schematic, the system has combined $\mathcal{C}_4$ and time-reversal symmetry with the rotation axis running through the center of the empty plaquettes. (b) Generalizing the concept of altermagnetism, materials breaking any $\mathbb{Z}_2$ symmetry (indicated by red or blue) while preserving their combination with rotational symmetry in the lattice, defines a type of altermaterial. In the example shown, rotating the lattice by $90$ degrees and inverting the colors leaves the figure unchanged. (c) Here, we introduce alterelectrics, where the $\mathbb{Z}_2$ symmetry is the symmetry flipping electric dipoles into and out of the plane.}
\end{center}
\end{figure}

%%%%%%%%%%%%
\emph{Introduction---}Altermagnets take a special place in the family of time-reversal-symmetry breaking states of matter, because they preserve a combination of spin rotations and discrete lattice rotation or mirror symmetries (Fig. \ref{fig1}a), rather than the combination of spin rotations with translation or inversion symmetry found in antiferromagnets~\cite{vsmejkal2020crystal, vsmejkal2022anomalous}. This combination allows for a range of uncommon physical properties inviting direct application. For example, despite not having a net magnetization, d-wave altermagnets have spin-split bands as well as a unique quadrupolar piezomagnetic response that can cause magnetization of opposite sign when applying strain in orthogonal ways~\cite{xu2025alterpiezoresponse, ma2021multifunctional}. Altermagnets with $d$-wave symmetry are moreover a natural place to look for hyperbolic saddle points, because their bulk spin-split bands necessarily come together at high-symmetry points in such a way that to leading order the two bands have dispersions of the form $ \pm(k_x^2 - k_y^2)$, with different signs for opposing spins. Such hyperbolic dispersions have long been studied in the context of hydrodynamic internal waves~\cite{staquet_internal_2002} and have been of recent interest in the context of photonic and phononic devices because of their unique wave-guiding capabilities~\cite{hyperbolic,hyperbolic2,hyperbolic3,gomez2016flatland,hu2020moire,yves_twist-induced_2024}. Their natural occurrence in altermagnets allows for similarly unique spin transport properties. Finally, because altermagnets break both time-reversal symmetry and the combination of time-reversal and inversion symmetry, they allow for an anomalous Hall effect in the absence of net magnetization~\cite{vsmejkal2022emerging, leiviska2024anisotropy, reichlova2024observation}. This opens the way to practical applications of topology~\cite{song2025altermagnets}, including protected edge state that may be used as robust qubits~\cite{ghorashi2024altermagnetic} and advanced spintronic components~\cite{gonzalez2021efficient}. 
%
%%%%%%%%%%%%%%%%%%%%%%%%%%
%%%%%%%%%%%%%%%%%%%%%%%%%%

Because all of these properties of altermagnets are based on symmetry only, one may readily generalize them to non-magnetic states of matter (Fig \ref{fig1}b). So far, such extensions have been based on the combination of time-reversal and rotation symmetry~\cite{alterphotonic}. Here we substitute time-reversal symmetry by spatial inversion, suggesting the electric dipole as the key building block, which we then configure to create \emph{alterelectric} materials (Fig. \ref{fig1}c). We find that some of the physical features of altermagnets---hyperbolic dispersion, anisotropic piezo-response, split bands, polarized transport---can be translated into the alterelectric setting. In stark contrast to altermagnets with bulk spin-polarized transport, however, there are no `dipole-split' bands in the alterelectric to carry a polarized bulk current. We therefore consider a finite-sized alterelectric material with surface states to introduce transport along distinct surfaces rather than distinct spin states. This opens a path towards ``surface-tronics'' as a spinless analog to spintronics.

\begin{figure}[t!]
\begin{center}
\includegraphics[width=0.86\columnwidth]{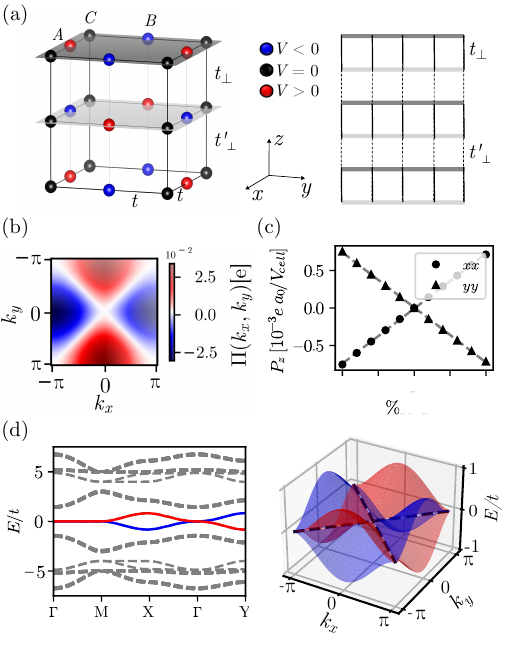}
\caption{\label{fig2} {\bf Multipolar piezoelectricity and hyperbolic surface modes in an alterelectric.} (a)	On the left: a unit cell of an ``altermagnet-like" model of dipoles, where each dipole consists of a pair of red (on-site potential $V>0$) and blue ($V<0$)  sites in a bilayer. On the right: a schematic side view of the bilayer structure, showing the alternating bond lengths and strengths along $z$.  
(b)	The partial polarization $\Pi(k_x, k_y)$ along the $z$ axis as calculated by the Berry phase method (see Eq. \eqref{eq:Pi}), displayed across the first Brillouin Zone.
(c)	The total polarization as a function of strain along the $x$ and $y$ directions.
(d)	The band structure of a slab finite in $z$ (20 unit cells), and infinite along $x$ and $y$. The gray dashed lines indicate the bulk bands. The state colored in red is localized on the top surface of the slab and that in blue on the bottom surface. The figure on the right displays the surface state dispersion across the first Brillouin zone, showing that these are two hyperbolic surfaces rotated $90^\circ$ with respect to one another. The parameter values used are $V= 4$, $t=1$, $t_\perp = 3$ and $t'_\perp =0.1$. For plotting, VESTA~\cite{vesta} and matplotlib~\cite{matplotlib} were used.}
\end{center}
\end{figure}

%%%%%%%%%%
\emph{Minimal model for an alterelectric---}We first introduce a minimal implementation of an alterelectic (Fig. \ref{fig2}a). It consists of bilayers of identical Lieb lattices that are rotated by a $90^\circ$ angle. In each Lieb lattice, the central (black) atom is neutral, whereas the two atoms on distinct links have opposite onsite potential $\pm V$, inspired by the minimal altermagnetic model in~\cite{brekke2023two}. This creates two dipoles of opposite direction along the $z$-axis in each unit cell. Crucially, flipping the dipoles is equivalent to rotating the lattice by $90^\circ$ or an inversion, and not to any translation. The dielectric analog to altermagnets is thus defined by a rotoinversion. We then create a stack of bilayers by alternating the interlayer bond strength in the third dimension. The tight-binding Hamiltonian for the system is given by:
\begin{align}
    H &= \sum_l \sum_{\textbf{k}} (-1)^l V \left( \hat{a}^{\dagger}_{l,\textbf{k}} \; \hat{a}^{\phantom\dagger}_{l,\textbf{k}} - \hat{b}^{\dagger}_{l,\textbf{k}}\; \hat{b}^{\phantom\dagger}_{l,\textbf{k}} \right) + \notag \\
    &2t \left[ \cos\left(\frac{k_xa}{2}\right) \hat{c}^{\dagger}_{l,\textbf{k}} \; \hat{a}^{\phantom\dagger}_{l,\textbf{k}} + \cos\left(\frac{k_ya}{2}\right) \hat{c}^{\dagger}_{l,\textbf{k}} \; \hat{b}^{\phantom\dagger}_{l,\textbf{k}} + \text{H.c.} \right] + \notag \\
    &\left(t_{z}+(-1)^l \delta_z \right) \left[ \hat{c}^{\dagger}_{l,\textbf{k}} \; \hat{c}^{\phantom\dagger}_{l+1,\textbf{k}} + \hat{a}^{\dagger}_{l,\textbf{k}} \; \hat{a}^{\phantom\dagger}_{l+1,\textbf{k}} \right. \notag \\
    & \phantom{- \left(t_{z}+(-1)^l \delta_z\right) ()} \left. +\;  \hat{b}^{\dagger}_{l,\textbf{k}} \; \hat{b}^{\phantom\dagger}_{l+1,\textbf{k}} + \text{H.c.}\right]
    \end{align}
Here, the operators $\hat{a}^\dagger$, $\hat{b}^\dagger$, and $\hat{c}^\dagger$ create electrons on the $A$, $B$, and $C$ sublattices indicated in Fig. \ref{fig2}a, while $a$ is the lattice parameter. The first two lines indicate the intra-layer Lieb-lattice model, while the final terms denote inter-layer hopping. The vector $\textbf{k}$ indicates in-plane momentum, while $l$ is a layer index defined such that each unit cell contains one layer with an even value of $l$ and one with an odd value. The Hamiltonian can alternatively be thought of as Su-Schrieffer-Heeger (C sublattice with black atoms) and Rice-Mele (A and B sublattices, with red and blue atoms) chains running in the $z$-direction~\cite{su1980soliton,rice1982elementary}, interacting through the in-plane coupling $t$. We choose values of the hopping parameters $t $, $t_\perp = t_z + \delta_z$, and $t'_\perp = t_z - \delta_z $ such that the band structure is gapped (insulating) at half filling (Fig. \ref{fig2}a,d). 
%
%%%%%%%%%%%%%%%%%%%%%%
%%%%%%%%%%%%%%%%%%%%%%

%%%%%%%%%%%%
\emph{Polarization and piezoelectricity---}The Berry phase of the occupied bands computed along $k_z$ (Fig. \ref{fig2}b) and integrated over $k_x$ and $k_y$ is proportional to the polarization along the $z$-direction~\cite{king1993theory, resta1994modern}. To be precise, the polarization along $z$ is given by:
\begin{align}
    P_z = \frac{e}{2 \pi V_{\text{cell}}} \int_{BZ} dk_x dk_y  \phi(k_x, k_y).
\end{align} 
To account for degeneracies in the occupied bands, the path along $k_z$ can be split into $N$ segments: 
\begin{align}
\phi(k_x, k_y) &\approx 2\pi \Pi(k_x, k_y) = - \text{Im}\left[\ln\left(\det\left(\Pi_{i = 0}^{N-1} M^{i, i+1}\right)\right)\right].
\label{eq:Pi}
\end{align} 
Here, $M^{i, i+1}_{m,n} = \langle u_{m, k_{z, i}}|  u_{n, k_{z, i+1}}\rangle$ indicate overlap matrix elements between bands $m$ and $n$. This approximates the exact result for a sufficiently dense k-point mesh~\cite{vanderbilt2018berry}. The symmetry exhibited by the partial polarization $\phi(k_x, k_y)$ is analogous to that of the magnetization $M(k)$ in altermagnetic materials: $\phi(k_x, k_y)$ maps onto itself by a combination of fourfold rotation and a flip of the polarization. By symmetry, the net polarization is guaranteed to be zero. This symmetry enforced vanishing of the total electric polarization in our alterelectric is equivalent to the symmetry enforced vanishing of the total magnetization in altermagnets.

Breaking the rotoinversion symmetry brings the next distinctive property of altermagnetism to alterelectrics---their response to deformation. The total electric polarization $P_z$ obtained by straining in one orientation is equal and opposite to that obtained by straining in the orthogonal direction (Fig \ref{fig2}c). Here, strain is implemented in the minimal model by assuming an exponential dependence of the hopping parameters on bond lengths (the codes used for the computations are available at Ref.~\cite{zenodo}).  Like in altermagnets, the anisotropic response is a bulk property. Unlike in altermagnets, however, it is not possible to define gauge-invariant local moments, and the net polarization in Fig. \ref{fig2}c is the manifestation of a bulk change in the electronic configuration. 

%%%%%%%%%%%%%%%%%%%%%%
\begin{figure*}[t!]
\begin{center}
\includegraphics[width=2\columnwidth]{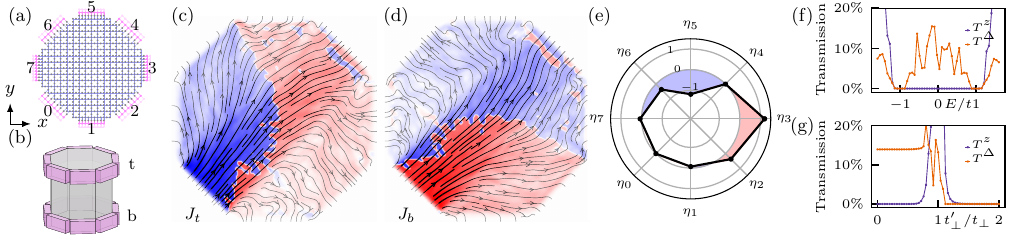}
\caption{\label{fig3} {\bf Anisotropic surface transport in an alterelectric.} (a) Top view of the octagonal transport setup with radius $R=8a$. The gray region is the scattering region and the numbered leads are in pink. (b) Schematic of the three-dimensional finite system used in the computations. There are 16 leads attached to the eight edges of the octagonal scattering region, with half of the leads at the top and half at the bottom. (c) and (d) show the top and bottom surface currents computed using states injected in lead zero at $E=0.2t$. Black arrows show the direction of the current while its magnitude is indicated by the intensity of the colormap. The red and blue colors indicate currents that are pointing more towards lead 2 or lead 6 respectively. The current on the top and bottom deviate in opposite directions from the $45\deg$ direction. (e) Radar plot of the ratio $\eta_i=10\log{T^{tt}_{i0}/T^{bb}_{i0}}$. The red (blue) area shows the direction in which the top (bottom) surface has larger transmission. (f) and (g) show the transmission probability as a function of scattering energy and as a function of the ratio $t'_\perp/t_\perp$ of the hopping amplitudes in the $z$ direction. The transmission in the $z$ direction is computed as $T^z = \sqrt{\sum_{i} |T^{tb}_{0i}|^2}/T_0$. The difference in transmission between the two surfaces is computed as $T^{\Delta}=\sqrt{\sum_{i} |T^{tt}_{0j}-T^{bb}_{0j}|^2}/T_0$. The $T_0$ normalization is chosen as $T^{tt}_{04}(E=0,t_\perp'=0.5t)$ in panel (f) and as $T^{tt}_{04}(E=0.2,t_\perp'=0)$ in panel (g).} 
\end{center}
\end{figure*}
%%%%%%%%%%%%%%%%%%%%%%

%%%%%%%%%%%%
\emph{Surface states---}The minimal model can be viewed not only as a collection of bilayers generating altermagnet-like dipole moments, but also as an assembly of coupled Rice-Mele chains running along the $z$-axis~\cite{rice1982elementary}. Such chains are known to have two zero-dimensional electronic states when the chain consists of an integer number of unit cells, localized on either end of the chain. We find that a finite slab of the alterelectric model may host states localized at the surfaces (Fig. \ref{fig2}d). These can be thought of as arising from the edge modes of individual Rice-Mele chains hybridizing under the in-plane coupling to form two-dimensional surface states. Because there is no symmetry pinning these surface states to zero energy, they can in principle be shifted into the bulk spectrum by increasing $V$. Nevertheless, the surface states exist for a wide of range of parameter values.

Importantly, the dispersion of the surface states seen in Fig. \ref{fig2}d reflects the rotoinversion symmetry of the model, which is also present in the two Lieb lattices constituting the crystal faces. The surface modes localize on opposite faces of the material, while their dispersions differ by a four-fold rotation. This implies that wave-packets propagating on opposite surfaces will be steered in different directions, yielding electronic transport properties with distinct anisotropies on each of the surfaces. This is analogous to the spintronic quality of altermagnets, in which currents with opposite spin-polarization have different anisotropic resistivities reflecting the symmetry of the spin-split bands~\cite{wu2024valley, gonzalez2021efficient}. Note that such spin-dependent anisotropy is a bulk property, whereas the transport of surface-localized wave packets in the alterelectric model is a boundary property, giving rise to ``surface-tronics''.

%%%%%%%%%%%%
\emph{Anisotropic surface transport---}Inspired by hyperbolic metamaterials, which exhibit hyperbolic bands and directional focusing~\cite{hyperbolic,hyperbolic2,hyperbolic3,gomez2016flatland, hu2020moire,yves_twist-induced_2024}, we demonstrate anisotropic surface transport by constructing a finite three-dimensional lattice with electronic contacts as shown in Fig.~\ref{fig3}ab. The octagonal column reflects the symmetry of the Hamiltonian and allows us to numerically establish the transmission probabilities between different contacts, as detailed in Appendix~\ref{app:transport}. The transmission probabilities between different contacts are computed using the scattering matrix and described by the matrix $T^{\alpha\beta}_{ij}$, where $\alpha,\beta\in\{t,b\}$ are surface indices and $i,j\in\{0, \dots, 7\}$ are the lead indices defined in Fig.~\ref{fig3}a. Because time-reversal symmetry is preserved, the transmission probabilities must obey the relation $T^{\alpha\beta}_{ij}=T^{\beta\alpha}_{ji}$.

The bulk dispersion is gapped at zero energy, so that $T^{tb}_{ij}\approx 0$ there. We therefore focus on transport across the top and bottom surfaces $T^{tt}_{ij}$ and $T^{bb}_{ij}$.
Figs.~\ref{fig3}c and \ref{fig3}d show that typical current profiles are asymmetric: in the top surface current injected from lead 0 is deflected towards lead 2, whereas in the bottom surface the current is deflected towards lead 6. The hyperbolic nature of the surface state dispersion thus results in anisotropic transport that could be used to a create anti-aligned voltages on opposing surfaces in the direction orthogonal to the current. To further quantify the anisotropy and difference between the two surfaces, we define $\eta_i=10\log{T^{tt}_{i0}/T^{bb}_{i0}}$. The regions with the highest values of $\eta_i$ being separated in Fig. \ref{fig3}e by a large angle suggests that surfacetronic transport could be used to split a current into two spatially separated currents with different directions, in analogy to the spin current splitting achieved by altermagnets. 

To further clarify the role of surface modes in the anisotropic transport properties, we show in Fig.~\ref{fig3}fg the transmission in the $z$ direction, defined as $T^z = \sqrt{\sum_{i} |T^{tb}_{0i}|^2}/T_0$. This transmission becomes significantly non-zero whenever the bulk band gap closes. We also display $T^\Delta = \sqrt{\sum_{i} |T^{tt}_{0i}-T^{bb}_{0i}|^2}/T_0$, which is indicative of the difference in anisotropic transport across the two surfaces. For both quantities, the normalization constant $T_0$ is chosen using $T^{tt}_{04}$ at convenient parameter values. Figure~\ref{fig3}f shows $T^{z}$ and $T^{\Delta}$ as a function of the energy $E$ of the incoming electrons. Around $E=0$, the system is gapped, as signaled by the vanishing $T^{z}$. Around $|E|=t$ the gap closes and bulk bands appear, as shown by the increase of $T^{z}$. At values of $E$ where the bulk is gapped, the surface states still conduct and transport along the surfaces is possible. The $T^{\Delta}$ difference between the two surfaces has a minimum at $E=0$, and is maximal halfway between $E=0$ and the edge of the band gap. The non-trivial shape of the peaks in $T^\Delta$ are due to finite-size effects and are expected to smoothen out for larger systems in the presence of some disorder.

Because the surface states originate from the edge modes of Rice-Mele chains, they can be removed by driving the model parameters through a topological phase transition. At this transition, the gap in the bulk spectrum closes, and transport between opposing surfaces becomes possible, as shown by the non-zero $T^z$ value in Fig.~\ref{fig3}g. In the trivial phase beyond the transition the system is fully gapped with vanishing $T^{\Delta}$ and no transmission between leads.

So far we have shown that a minimal realization of an alterelectric yields an array of model properties that are analogous to those of altermagnets. We next show that the symmetries of the minimal model of Fig.~\ref{fig2} can also be used as a guiding principle for engineering first-principle models of realistic materials that show the piezoelectric behaviour of $d$-wave alterelectrics.

%%%%%%%%%%%%%%
\emph{Ab initio implementation---}Realizations of simple models in bulk materials are generally scarce, primarily due to the presence of ubiquitous additional couplings and/or interactions, and the abundance of low-symmetry structures. In particular, no viable candidates for Lieb-lattice materials are known to date. The closely related checkerboard lattice also known as the inverse Lieb lattice, however, closely resembles the layers realized in perovskite-like structures.

\begin{figure}[t!]
    \includegraphics[width=8.6cm]{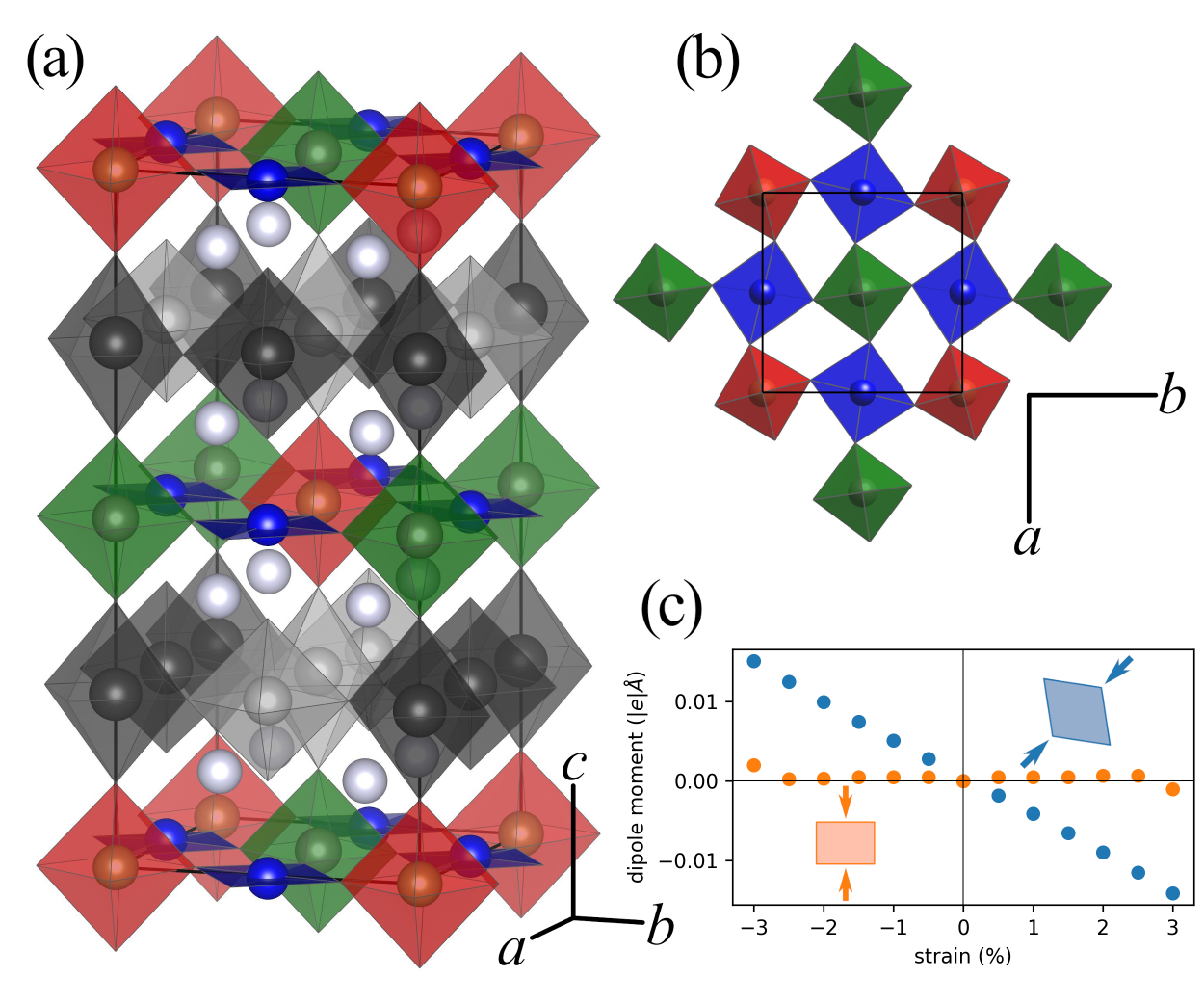}
    \caption{\label{fig:dft} {\bf First-principles model for an alterelectric material.} (a) The $P\bar{4}$ unit cell of the fictitious, fully ordered Sr$_4$CuTe$_{0.5}$W$_{0.5}$TiZrO$_{12}$ material. The building blocks are CuO$_4$ plaquettes (blue) and four kinds of octahedra: TeO$_6$ (red), WO$_6$ (green), TiO$_6$ (gray), and TiO$_6$ (dark gray). Sr atoms are light gray spheres. (b) Top view of a magnetic layer showing the alternation of TeO$_6$ octahedra. (c) Calculated strain-induced dipole moment ($z$-component) for pressure applied along $a$ (orange) and the $ab$ diagonal direction (blue). Negative strain values correspond to compression along $b$ and the other diagonal instead. For plotting, VESTA~\cite{vesta} and matplotlib~\cite{matplotlib} were used.}
\end{figure}

This is exemplified by Sr$_2$Cu$X$O$_6$ with $X$ = Te or W~\cite{koga16,vasala14}. Both are magnetic insulators with the double perovskite structure. The $x^2-y^2$ magnetically active orbitals of Cu$^{2+}$ gives rise to strongly quasi-2D magnetism described by the frustrated square lattice model, and a possible spin liquid state in Te/W disordered realizations~\cite{watanabe18, mustonen18, mustonen23}. Here, we take an alternative route by considering the fictitious, fully ordered material Sr$_2$CuTe$_{0.5}$W$_{0.5}$O$_6$ where Te and W alternate in the magnetic layers (Fig.~\ref{fig:dft}b). Such alternation is expected to systematically weaken second-neighbor bonds in every second square of the lattice, thereby leaving the equivalence of first-neighbor bonds. The resulting exchange interactions are described by a lattice which can be thought of as an interpolation between the original frustrated square lattice and the checkerboard lattice. The latter component is crucial, as its symmetry would give rise to altermagnetism in a two-sublattice magnetic state. Since we are interested in alterferroelectricity rather than altermagnetism, however, we consider only uniform (ferro)magnetic configurations, with trivial (but asymmetric) spin splittings.

Next, we need to remove mirror symmetries that connect neighboring magnetic layers to avoid ending up with an antiferroelectric. To this end, we double the unit cell and additionally introduce two nonmagnetic perovskite spacer layers, with nonmagnetic cations Ti and Zr. Both can be $4+$, as required for electroneutrality, and each atomic species is included in stripes running along the $a$ ($b$) axis in the first (second) spacer layer. The resulting
Sr$_4$CuTe$_{0.5}$W$_{0.5}$TiZrO$_{12}$ structure (Fig.~\ref{fig:dft}a) has a unit cell with 80 atoms, features four perovskite-like layers and conforms to the desired space group $P\bar{4}$. To turn this prototype into a fictitious yet realistic structure, we optimize the internal atomic coordinates and the $c$ lattice parameter, by fixing the in-plane lattice constant to the average value of bulk Sr$_2$CuTeO$_6$ and Sr$_2$CuWO$_6$. 

With the optimized structural parameters, we calculate the electric polarization using the Berry-phase method implemented in vasp version 6.4.2~\cite{resta1994modern, vasp, vasp_2, vasp_pseudo}. As the $\bar{4}$ point group is not polar, the original structure features zero magnetization and polarization. To study its piezoelectric response, we estimate the effect of applying uniaxial pressure either along the coordinate directions $a\equiv[100]$ and $b\equiv[010]$, or along the diagonals $[110]$ and $[1\bar{1}0]$. This is done keeping the volume constant by compensating longitudinal squeezing with transverse
expansion while assuming that the fractional atomic coordinates remain fixed. Under these conditions, we observe a linear strain dependence of polarization along $[001]$ if pressure is applied along the diagonal (blue circles in Fig.~\ref{fig:dft}c). In contrast, the electric polarization resulting from pressure in coordinate directions does not
exceed the numerical error (orange circles in Fig.~\ref{fig:dft}c).

While the atomic structure considered here is a fictitious material introduced only to illustrate the possibility of designing materials guided by alterelectric symmetry constraints, we note that perovskite films and heterostructures thereof are in fact routinely grown on various substrates and in ultra-thin films~\cite{cheng2022fabrication, park2023two}.

%%%%%%%%%%%%
\emph{Conclusions and outlook---}By constructing altermagnet-like sublattices containing elements with opposite $\mathbb{Z}_2$ charges of any kind, properties of altermagnets can be explored without breaking time-reversal symmetry. We demonstrated this for the case of electric polarization using a minimal model displaying $d$-wave polarization and anisotropic piezoelectric response, as well as surface states with hyperbolic dispersion and corresponding anisotropic surface transport allowing for surface-split currents and surfacetronic applications. All of these are direct analogs of characteristic altermagnetic properties in the alterelectric stetting. We also showed that the symmetries of the alterelectric model can be used to guide the design of a first principles model embodying a fictitious but realistic alterelectric atomic and electronic structure.

The symmetries and materials properties introduced here may be implemented across a broad class of systems, including topological ones, where the increased robustness of surface states would enhance their suitability for applications. The analogy to altermagnets utilized in this work is not limited to systems based on electric polarization, but may be extended to any system that breaks a $\mathbb{Z}_2$-symmetry, either statically or dynamically, providing a broad arena for novel (meta)material design.

%%%%%%%%%%%%
\emph{Note added during editing---}In the final stages of preparing this manuscript, a preprint of an independent study of alterelectrics appeared~\cite{preprint26}. As in the alterelectric structures defined here, they focus on systems in which a pair of switchable electronic states with distinct band structures are related by a symmetry other than translation, in such a way that the switching and symmetry operations have the same effect. In contrast to the discussion of anisotropic piezoelectricity, surface transport, and realization in bulk crystals presented here, however, they focus on the construction of an electric quadrupole tensor as the Landau order parameter, and its possible realization in sliding bilayers and ionically adsorbed monolayers.

%%%%%%%%%%%%
\emph{Acknowledgments---}The authors would like to thank Sander Mann for stimulating discussions throughout the research work, and Ulrike Nitzsche for technical assistance. C.C. acknowledges funding from the European Research Council (Grant Agreement ERC-CoG 101170693) and from the Netherlands Organisation for Scientific Research (Grant Agreement No. VIDI 2131313). J.v.d.B. and O.J. acknowledge support from the German Forschungsgemeinschaft (DFG, German Research Foundation) through SFB 1143 (Project ID 247310070). V.K. was funded by the European Union. A.C.H.V. acknowledges support by the Huo Family Foundation through a P.C. Ho PhD Studentship.

%%%%%%%%%%%%
\emph{Author contributions---}The project was conceived by C.C. and J.v.W., and developed with input from A.C.H.V., V.K., and J.v.d.B.. The theoretical checkerboard model was introduced by J.v.d.B. and refined into the Lieb lattice model by V.K.. Numerical calculations of the polarization and surface states were performed by A.C.H.V. under supervision of V.K. and J.v.W.. V.K. carried out the numerical transport calculations. The material realization was proposed and numerically investigated by O.J.. All authors contributed to the interpretation of the results and to the writing of the manuscript.

%%%%%%%%%%%%
%

%%%%%%%%%%%%
\appendix
\section{Transport calculations}
\label{app:transport}
We consider a multi-terminal transport setup, in which leads are attached to all sides on the top and bottom of the octagonal column scattering region shown in Fig.~\ref{fig3}ab. The Kwant package is used to calculate the resulting scattering matrix and the current densities associated with the scattering states~\cite{Groth2014}. The codes used for the computations are available at Ref.~\cite{zenodo}.

For the leads, an orthorhombic lattice was attached to the top and bottom four layers. The onsite energy in the leads is set to match the scattering energy, which ensures that at any scattering energy, the same propagating modes are addressed in the leads. We thus probe energy variations of the system rather than the leads. 

Throughout the calculations, the parameters used were $V=4t$, $t_\perp=3t$, and $t_\perp'=0.5t$.

%%%%%%%%%%%%
\section{Alterelectric Checkerboard Lattice}
\label{app:checkerboard}
\begin{figure}[b]
\begin{center}
\includegraphics[width=0.86\columnwidth]{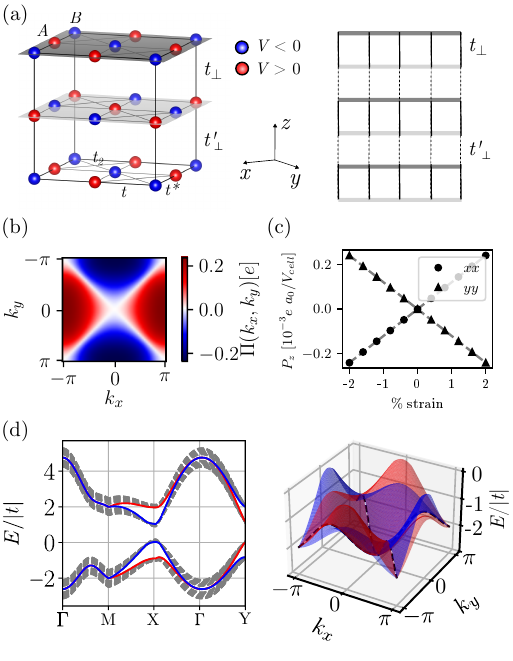}
\caption{\label{checkerboard} {\bf Multipolar piezoelectricity and surface modes in the alterelectric checkerboard lattice} (a)	On the left: a unit cell of an ``altermagnet-like" model of dipoles, where each dipole consists of a pair of red (on-site potential $V>0$) and blue ($V<0$)  sites in a bilayer. On the right: a schematic side view of the bilayer structure, showing the alternating bond lengths and strengths along $z$.  
(b)	The partial polarization $\Pi(k_x, k_y)$ along the $z$ axis as calculated by the Berry phase method (see Eq. \eqref{eq:Pi}), displayed across the first Brillouin Zone.
(c)	The total polarization as a function of strain along the $x$ and $y$ directions.
(d)	The band structure of a slab finite in $z$ (20 unit cells), and infinite along $x$ and $y$. The gray dashed lines indicate the bulk bands. The states colored in red are localized on the top surface of the slab and those in blue on the bottom surface. The figure on the right displays the dispersion of the two lower energy surface states across the first Brillouin zone, showing that these are two hyperbolic surfaces rotated $90^\circ$ with respect to one another. The parameter values used are $V= 1$, $t=1+0.5i$, $t2 = 0.6$, $t_\perp = 0.4$ and $t'_\perp =0.2$.}
\end{center}
\end{figure}

To illustrate that the polarization properties and edge modes discussed in the main text are not limited to the Lieb lattice, we here consider a model based on checkerboard planes (Fig. \ref{checkerboard}a), as inspired by Ref.~\cite{yershov2024fluctuation}. The lattice is again constructed by stacking layers related by a 90 degree rotation (or equivalently by exchanging the on-site potentials) and stacking bilayers connected by alternating hopping parameters. To open a gap between the bands, as required for obtaining a polarization, a complex NN hopping $t$ respecting the rotoinversion symmetry of the system is introduced. The tight-binding Hamiltonian for this system is then given by: 
\begin{align}
    H &= \sum_l \sum_{\textbf{k}} (-1)^l V \left( \hat{a}^{\dagger}_{l,\textbf{k}} \; \hat{a}^{\phantom\dagger}_{l,\textbf{k}} - \hat{b}^{\dagger}_{l,\textbf{k}}\; \hat{b}^{\phantom\dagger}_{l,\textbf{k}} \right) +\notag \\& 2t_2\left(\cos(k_xa)\hat{a}^{\dagger}_{l,\textbf{k}} \; \hat{a}^{\phantom\dagger}_{l,\textbf{k}} + \cos(k_ya)\hat{b}^{\dagger}_{l,\textbf{k}} \; \hat{b}^{\phantom\dagger}_{l,\textbf{k}}\right) +\notag  \\
    &2\left[ t \cos\left(\frac{k_x+k_y}{2}a\right) \hat{a}^{\dagger}_{l,\textbf{k}} \; \hat{b}^{\phantom\dagger}_{l,\textbf{k}} +\notag \right.\\ &\left.t^* \cos\left(\frac{k_x-k_y}{2}a\right) \hat{a}^{\dagger}_{l,\textbf{k}} \; \hat{b}^{\phantom\dagger}_{l,\textbf{k}} + \text{H.c.} \right] + \notag \\
    &\left(t_{z}+(-1)^l \delta_z \right) \left[ \hat{a}^{\dagger}_{l,\textbf{k}} \; \hat{a}^{\phantom\dagger}_{l+1,\textbf{k}} \right. \notag \\
    & \phantom{- \left(t_{z}+(-1)^l \delta_z\right) ()} \left. +\;  \hat{b}^{\dagger}_{l,\textbf{k}} \; \hat{b}^{\phantom\dagger}_{l+1,\textbf{k}} + \text{H.c.}\right]
\end{align}

Here operators $\hat{a}^{\dagger}_{l,\textbf{k}}$ and $\hat{b}^{\dagger}_{l,\textbf{k}}$ create electrons on the $A$ and $B$ sublattices indicated in Fig.~\ref{checkerboard}a. The first four lines indicate the intra-layer checkerboard lattice model, while the remaining terms indicate inter-layer hopping. As in the main text, $\textbf{k}$ indicates in-plane momentum and $l$ is a layer index. 

The Berry phase $\phi(k_x, k_y)$ is defined as in Eq.~\eqref{eq:Pi} and displayed in Fig.~\ref{checkerboard}b. It exhibits the same symmetry and analogy to the altermagnetic $M(k)$ as in the Lieb model discussed in the main text. Consequently, the same type of quadrupolar piezoelectric response is present, as shown in Fig.~\ref{checkerboard}c. A finite section of this model, which can again be though of also as parallel Rice-Mele chains, also exhibits surface modes (Fig.~\ref{checkerboard}d). However, unlike in the Lieb model, these occur in gaps within the two lowest and the two highest energy bands, rather than around zero energy. Zooming in onto the two lowest-energy surface states on the right of Fig.~\ref{checkerboard}d shows them to be related by a 90 degree rotation, completing the analogy to the results discussed in the main text.

%%%%%%%%%%%%
\section{Band-structure calculations}\label{app:dft}
To construct the initial prototype of Sr$_4$CuTe$_{0.5}$W$_{0.5}$TiZrO$_{12}$, we took the average of the $I4/m$ crystal structures of Sr$_2$CuTeO$_6$~\cite{koga16} and Sr$_2$CuWO$_6$~\cite{vasala14}, defined a $\sqrt{2}\times\sqrt{2}\times{}1$ supercell, interleaved the magnetic layers with nonmagnetic Sr$_2$TiZrO$_6$ layers, and doubled the $c$ parameter to accommodate to the doubled number of layers. The as-constructed structure is described by the $P\bar{4}$ space group with $a=b=7.65768$ \r{A} and $c_0=16.8646$ \r{A}.

In the next step, we constructed several prototype structures featuring the same in-plane lattice parameters and internal atomic coordinates, but different $c$ values ($c=c_0\pm1$ \r{A}). For each structure, we performed density-functional theory (DFT) calculations using the Perdew-Burke-Ernzerhof (PBE) functional~\cite{PBE96} and optimized the internal coordinates with respect to the total energy. In these calculations, we used a 400 eV energy cutoff for the plane-wave basis and a $k$-mesh of $4\times{}4\times{}2$ points. Electronic correlations in the partially filled $3d$ shell of Cu were treated
using the rotationally invariant version of DFT+$U$ with the Coulomb repulsion $U_d$ = 9.5 eV and the Hund exchange $J_d$ = 1 eV; initial moments of all Cu atoms were set to 1 $\mu_{\text{B}}$. In this way, we found the minimal total energy for the structure depicted in Fig.~\ref{fig:dft}a with the lattice parameter $c=16.8946$ \r{A}. The resulting Hellmann–Feynman forces do not exceed 0.0015 eV/\r{A}.

Electric polarization calculations were performed using a 400 eV energy cutoff and a $k$-mesh of $8\times{}8\times{}4$ points.

%%%%%%%%%%%%
\end{document}